\documentclass[prb,aps,twocolumn,amsmath,amssymb,showpacs]{revtex4}
\usepackage{graphicx}
\usepackage{amssymb,amsmath}

\begin{document}

\title{      Microscopic  origin of diagonal stripe phases  
             in doped nickelates}

\author{     Marcin Raczkowski}
\affiliation{Marian Smoluchowski Institute of Physics, Jagellonian
             University, Reymonta 4, PL-30059 Krak\'ow, Poland}
\affiliation{Laboratoire CRISMAT, UMR 6508 CNRS-ENSICAEN, 6, Boulevard
             du Mar\'echal Juin, 14050 CAEN Cedex, France}

\author{     Raymond Fr\'esard}
\affiliation{Laboratoire CRISMAT, UMR 6508 CNRS-ENSICAEN, 6, Boulevard
             du Mar\'echal Juin, 14050 CAEN Cedex, France} 

\author{     Andrzej M. Ole\'s}
\affiliation{Marian Smoluchowski Institute of Physics, Jagellonian
             University, Reymonta 4, PL-30059 Krak\'ow, Poland}

\date{\today}

\begin{abstract}
We investigate the electron density distribution and the stability of 
stripe phases in the realistic two-band model with hopping elements 
between $e_g$ orbitals at Ni sites on the square lattice, and compare 
these results with those obtained for the doubly degenerate Hubbard 
model with two equivalent orbitals and diagonal hopping. For both models 
we determine the stability regions of filled and half-filled stripe 
phases for increasing hole doping $x=2-n$ in the range of $x<0.4$, 
using Hartree-Fock approximation for large clusters. In the parameter 
range relevant to the nickelates, we obtain the most stable diagonal 
stripe structures with filling of nearly one hole per atom, as observed 
experimentally. In contrast, for the doubly degenerate Hubbard model 
the most stable stripes are somewhat reminiscent of the cuprates, 
with half-filled atoms at the domain wall sites. 
This difference elucidates the crucial role of the off-diagonal $e_g$ 
hopping terms for the stripe formation in La$_{2-x}$Sr$_x$NiO$_4$. 
The influence of crystal field is discussed as well.
\end{abstract}


\maketitle

\section{Stripe phases in nickelates}
\label{Sec:spn}

Stripe phases are one of the most exciting phenomena of modern
condensed matter physics. They have been observed in a variety of
systems, including nickelates,
\cite{Sac95,Tra96,Lee97,Yos00,Kaj01,Lee02,Fre02,Kaj03} cuprates, 
\cite{Tra95Nd,Tra96Nd,Tra97Nd,Yam98,Ich00,Sin99,Wak03} and manganites.
\cite{Dag03,Kim02,Lar05} Among them, layered La$_{2-x}$Sr$_x$CuO$_4$ 
(LSCO), La$_{2-x-y}$Nd$_{y}$Sr$_x$CuO$_4$ (Nd-LSCO), 
La$_{2-x}$Sr$_x$NiO$_4$ (LSNO), and La$_{2}$NiO$_{4+\delta}$ (LNO) 
compounds play plausibly the most 
prominent role. However the similarity between them is superficial only, 
and the stripes in the cuprates differ from the stripes in the 
nickelates in many respects. For instance, they are dynamical in the 
former, and static in the latter. In addition, in Nd-LSCO 
\cite{Tra95Nd,Tra96Nd,Tra97Nd,Ich00,Sin99,Wak03} and LSCO,\cite{Yam98} 
one finds the so-called {\it half-filled stripes\/}, with the density 
of one doped hole per two atoms along the domain wall (DW). In contrast, 
it is clear from a variety of experiments that magnetic states within 
doped NiO$_2$ planes of the nickelates are {\it filled stripes\/} with 
density of one doped hole per one atom in a DW.
\cite{Sac95,Tra96,Lee97,Yos00,Kaj01,Lee02,Fre02,Kaj03}  

The question of filling is not the only difference between the nickelate 
and cuprate stripes, however. Neutron diffraction measurements performed 
on Nd-LSCO revealed that magnetic peaks are displaced from the 
antiferromagnetic (AF) maximum at 
${\bf Q}_{\rm AF}=\pi(1,1)$ to the points 
${\bf Q}_{s}^{}=\pi(1\pm2\epsilon,1)$ and 
${\bf Q}_{s}^{}=\pi(1,1\pm2\epsilon)$ and the shift $\epsilon$ depends 
linearly on hole doping $x$ for $x<1/8$, while it is 
almost constant at higher doping. These values correspond to a 
superposition of \emph{vertical} (01) and \emph{horizontal} (10) DWs. 
The essentially identical modulation and doping dependence of $\epsilon$ 
was observed in superconducting crystals of LSCO with $x>0.05$.
Conversely, experiments on LSNO established that spin order is 
characterized by the wave vectors 
${\bf Q}_{s}^{}=\pi(1\pm\epsilon,1\pm\epsilon)$ with $\epsilon\simeq x$ 
for $x<1/3$, corresponding to a constant charge of one hole/Ni ion along 
a \emph{diagonal} DW, in agreement with the predictions made in the 
pioneering works by Zaanen and Gunnarsson\cite{Zaa89} and others,
\cite{Poi89,Sch89,Kat90,Inu91} and emphasized recently,\cite{Rac06} 
which predicted theoretically the stripe order. 
More precisely, neutron scattering measurements have revealed static 
stripe order in the LNO samples with $\delta=0.105$, 0.125, as well as 
0.133,\cite{Tra94O,Tra94Oa,Lor95,Tra95O,Tra97O,Tra97Oa,Woc98} and even 
over a wider hole doping regime $0.135\le x\le0.5$ in the case of LSNO.
\cite{Sac95,Tra96,Lee97,Yos00,Kaj01,Lee02,Fre02,Kaj03} Moreover, the 
incommensurate (IC) stripe order persists up to $x=0.7$ in the 
Nd$_{2-x}$Sr$_{x}$NiO$_4$ system,\cite{Ish03} in which the low
temperature orthorhombic phase seems to extend to a higher doping region 
$x\le 0.45$ as compared to the La compounds, where the high temperature 
tetragonal phase is stabilized instead already at $x\simeq 0.22$.
\cite{Sac95} 

Indications of a charge order (CO) in doped LSNO were also found in 
electron \cite{Che93} and x-ray diffraction studies.
\cite{Isa94,Vig97,Ish04} Quite recently, the one-dimensional nature of 
the stripe modulation within NiO$_2$ planes has been directly confirmed 
in the transmission electron microscopy (TEM) studies of charge stripes 
in LSNO.\cite{Li03} In addition, careful examination of the TEM images 
has shown that at a low temperature stripes are mainly centered on rows 
of Ni atoms. However, a mixture of the so-called site centered (SC) and 
bond centered (BC) stripes was also observed in some small regions of 
the sample. In contrast to LSCO, charge and spin order in LSNO is 
characterized by the wave vectors ${\bf Q}_{c}=2\pi(\epsilon,\epsilon)$ 
and ${\bf Q}_{s}=\pi(1\pm\epsilon,1\pm\epsilon)$, see Ref. 
\onlinecite{noteq}, respectively, with $\epsilon\simeq x$ corresponding 
to a constant charge density of one hole/Ni ion along the diagonal 
stripe. Note that the doping dependence of $\epsilon$ is exactly the 
same as that found in the seminal Hartree-Fock (HF) calculations within 
the Hubbard model.\cite{Zaa89,Poi89,Sch89,Kat90,Inu91}

Experimentally the data give a clear evidence that the stripe order is 
most stable at hole doping $x=1/3$. Indeed, $T_{\rm CO}^{\rm IC}$ and
$T_{\rm N}$ increase linearly with $x$, reach a maximum at $x=1/3$ with 
240 and 180 K, respectively, and then decrease monotonously upon further 
doping. The particularly robust stability of stripes at $x=\epsilon=1/3$ 
stems from the cooperative spin and charge modulation expressed by the
coincidence of the spin superlattice peaks at the wave vectors 
$(1/2\pm\epsilon/2,1/2\pm\epsilon/2)$ with those of CO, given by 
$(\epsilon,\epsilon)$. One should also note that $\epsilon$ starts to 
deviate gradually from the value given by the $\epsilon\simeq x$ law 
above $x=1/3$. In fact, CO is even more stable than the stripe 
order at $x=1/2$ and forms a checkerboard pattern below the transition 
temperature $T_{\rm CO}^{\rm C}\simeq 480$ K. Remarkably, with 
decreasing temperature, a stripe CO sets in below 
$T_{\rm CO}^{\rm IC}\simeq 180$ K and its incommensurability is twice 
as large as that of the spin order with the much lower onset temperature 
$T^{}_{N}\simeq 80$ K.\cite{Fre02,Kaj03} The low temperature competition 
of the checkerboard and stripe order at $x=1/2$ has also been clearly 
indicated by the measurements of Raman and optical conductivity.
\cite{Jun01,Yama03,Poirot02} Interestingly, above this doping,  
incommensurability tends to saturate rapidly with the value 
$\epsilon\simeq 0.44$.\cite{Ish03} 

Apparently, CO itself induces commensurate values of $\epsilon$ and such 
a commensurability effect seems to be an intrinsic property of the stripe 
order. At the same time, the increased hole density at the DWs 
suppresses the superexchange energy gain, best optimized when all the 
holes are accommodated within stripes. Therefore, it is the AF order that 
drives $\epsilon$ to approach the value given by the linear relation 
$\epsilon\simeq x$ below $T_{\rm N}$. Obviously, when $x=1/3$, both 
effects cooperate which results in the locked-in value of 
$\epsilon\simeq 1/3$ over the entire $T<T_{\rm CO}^{\rm IC}$ range.   

Furthermore, the temperature dependence of the specific heat $C_V$ shows 
a distinct anomaly at the same temperature $T_{\rm CO}^{\rm IC}=240$ K 
suggesting a CO transition.\cite{Ram96} This conjecture is supported 
by the temperature dependence of the logarithmic resistivity. Indeed,
as expected for this transition, $\log\rho$ of the sample with $x=0.3$ 
exhibits a steep increase precisely at 240 K.\cite{Che94} Charge 
fluctuations around $T_{\rm CO}^{\rm IC}=240$ K also lead to conspicuous 
changes in the optical conductivity spectra $\sigma(\omega)$.\cite{Kat96} 
Namely, at a high temperature $T>T_{\rm CO}^{\rm IC}$, only a broad peak 
is observed with a finite low-energy spectral weight. However, when $T$ 
is decreased down to $T_{\rm CO}^{\rm IC}$, the low-energy weight below
0.4 eV is gradually transferred to higher energy so that the opening of 
the charge gap is clearly observed. 

A special character of $x=1/3$ as well as $x=1/2$ doping, is best seen in 
the magnetic susceptibility $\chi$ and logarithmic resistivity 
$\log\rho$, recorded at 200 K, showing distinct anomalies at these doping 
levels.\cite{Che94} In this context it is important to discuss a peculiar 
behavior of the Hall coefficient $R_H$ and the Seebeck coefficient $S$ 
at two representative values of temperature below and above 
$T_{\rm CO}^{\rm IC}=240$ K, i.e., at 300 K and 210 K. The Hall 
coefficient at 300 K is almost independent of doping and takes a small 
positive value corresponding to the order of one hole carrier per Ni 
site.\cite{Kat99} The Seebeck coefficient shows a similar nearly 
constant behavior taking, however, a negative value. In contrast, below 
$T_{\rm CO}^{\rm IC}$, both $R_H$ and $S$ change their signs from 
negative to positive at $x=1/3$ and their absolute values are larger 
than those at 300 K. In addition, for samples with $x=0.3$ and $0.33$, 
$R_H$ keeps decreasing with decreasing $T$, so that the number of 
carriers per Ni site gets reduced even down to $0.01$.

These results indicate that the deviation of $x$ from $1/3$ can be 
considered as an electronlike ($x<1/3$) or holelike ($x>1/3$) carrier 
doping into the $\epsilon=1/3$ charge-ordered insulator with three Ni 
sites in the unit cell. Hence, for the  doping level $x=1/3$ there is 
exactly one hole per unit cell and such a state is robust and may be 
considered as a half-filled one. Moreover, it would certainly retain 
this feature if the incommensurability had followed precisely the 
relation $\epsilon=x$. However, $\epsilon$ has a tendency to shift 
towards $1/3$, for both sides of the $x=1/3$ point, which has important 
implications for the sign of $R_H$. On the one hand, when $x$ is less 
than $1/3$, the number of holes is insufficient for filling up the 
mid-gap states entirely, i.e., the states inside the charge transfer 
(CT) gap induced by stripe order, and in this case the mid-gap states 
contain some electrons which become carriers. On the other hand, for $x$ 
larger than $1/3$, the number of electrons is insufficient for filling 
up the lower Hubbard band entirely, which initially contains holes. 
Consequently, $R_H$ is expected to have the opposite sign to that in 
the case of $x<1/3$ (hole carriers). 
 
Next, the chemical potential shift $\Delta\mu$ in LSNO for $x\le 1/3$ 
deduced either from x-ray (XPS) or ultraviolet photoemission spectroscopy 
\cite{Sat00} is suppressed. Certainly, 
this phenomenon cannot be explained within a simple rigid-band framework 
in which $\mu$ is expected to shift downwards with increasing hole 
doping. In fact, an increase of $x$ in a system with a spatially uniform 
hole distribution should enhance the average hole-hole repulsion which, 
in turn, would result in a higher energy required to add one hole to the 
system, i.e., in a larger $|\mu|$. Therefore, the absence of $\Delta\mu$ 
implies that the average hole-hole repulsion remains nearly unaltered 
upon doping. Such a behavior might be easily explained within a stripe 
picture in which a constant hole density at the DWs implies that the 
interwall distance decreases and $\epsilon$ increases linearly with 
increasing $x$. Moreover, a similar suppression of the shift has been 
found below $x\le 1/8$ in LSCO, suggesting that an inhomogeneous charge 
distribution is a common feature of both systems.  

Several methods have been employed to investigate the stripe phases 
which go beyond the HF approximation, such as: 
density matrix renormalization group,\cite{Whi98,Whi99}
slave-boson approximation,\cite{Sei98,Sei98V,Sei04} 
variational local ansatz approximation,\cite{Gor99} 
exact diagonalization (ED) of finite clusters,\cite{Toh99} 
analytical approach based on variational trial wave functions within 
the string picture,\cite{Wro00,Wro05}
dynamical mean field theory,\cite{Fle00}   
cluster perturbation theory,\cite{Zac00} and 
quantum Monte Carlo simulations.\cite{Bec01,Rie01}  
They all address the crucial role of a proper treatment of local 
electron correlations in stabilization of the half-filled stripe phases 
in the cuprates. In spite of this huge effort, it remains unclear 
whether DWs are centered on rows of metal atoms as in  
SC stripes, or if they are centered on rows of oxygen atoms bridging 
two metal sites in BC stripes. 

Although no evidence was presented yet, it seems that the degeneracy of 
$3d$ orbitals plays an important role in stabilizing filled stripes in 
the nickelates. In the simplest picture developed for the cuprates, the 
Cu$^{3+}$ ions forming DWs are spinless, while the Cu$^{2+}$ of the AF 
domains carry a spin $S=1/2$. In the nickelates, filled DWs are formed 
of the Ni$^{3+}$ ions ($S=1/2$), whereas the AF domains consist of 
Ni$^{2+}$ ions ($S=1$). Therefore, a realistic Hamiltonian for LSNO has 
to contain, besides the $x^2-y^2$ orbital which is occupied by 
one hole in the parent compounds of the superconducting cuprates 
(such as La$_2$CuO$_4$ and YBa$_2$Cu$_3$O$_6$), as well the $3z^2-r^2$ 
orbital at each ion, so as to account for the high spin ($S=1$) state of 
the stoichiometric compound. Indeed, when extending their approach to a 
more realistic four-band Peierls-Hubbard model for NiO$_2$ planes, 
Zaanen and Littlewood have shown\cite{Zaa94} that doped holes prone to 
form diagonal DWs, centered on rows of Ni atoms, with a tendency to have 
a ferromagnetic (FM) alignment of the reduced Ni magnetic moments at a 
DW. In fact, subsequent multiband HF calculations emphasized the 
relevance of the electron-lattice coupling --- depending on its strength 
one can obtain either metal- or oxygen-centered structures.
\cite{Miz97,Yi98} Therefore, it appears that the semiclassical theory 
captures the essence of stripe physics in the nickelates.

The stripes were also obtained in the theory using either the HF 
approximation applied to a multiband CT model, in 
which both the nickel and oxygen degrees of freedom are explicitely 
taken into account,\cite{Zaa94,Miz97,Yi98} or by the ED of finite 
clusters within the effective two-band model (\ref{eq:H_deg}) extended 
by the coupling of $e_g$ electrons to the lattice.\cite{Hot04} However, 
due to a large number of basis states, the latter calculation has solely 
been done for an eight site cluster allowing only for investigating 
stripe phases with a small unit cell observed experimentally at high 
doping levels $x=1/3$ and $x=1/2$. In contrast, the HF approximation 
used here allows one to describe charge and spin modulation with larger 
unit cells and hence it should provide an answer to the important 
question whether the description of NiO$_2$ planes by the simplified 
effective $e_g$ model described below yields results consistent with 
the predictions of the CT model.\cite{Zaa94,Miz97,Yi98}
An important aspect of these studies, qualitatively different from the 
pioneering work on the cuprates where the relevant models yield DWs with 
\emph{nonmagnetic} Cu$^{3+}$ ions,\cite{Zaa89,Poi89,Sch89,Kat90,Inu91}
is that DWs in the nickelates are formed of Ni$^{3+}$ ions carrying 
a \emph{finite} spin $S=1/2$ leading to a FM polarization around them.
 
In this work we investigate various stripe phases resulting from an 
effective Hamiltonian for $e_g$ electrons relevant to the NiO$_2$ 
layers, and compare them with those obtained in the doubly degenerate 
Hubbard (DDH) model. We use the HF approximation in order to gain the 
qualitative insight into possible stripe phases and their stability. 
This approach is the first step to make in order to identify possible 
generic instabilities towards stripe phases in NiO$_2$ planes, in analogy 
to the early work on CuO$_2$ planes which predicted the existence of 
stripes in the high $T_c$ cuprates.\cite{Zaa89} The HF method is well 
suited to compare the energies of different types of magnetically 
ordered phases, as it approaches the same limit at $U\to\infty$ as 
better mean-field theories, such as slave-boson approach or the 
Gutzwiller ansatz.\cite{Ole89} In addition, the HF 
approach allows one for efficient calculations on large clusters at low 
temperatures which are necessary to obtain unbiased results concerning 
the stability of various stripe phases, particularly when the magnetic 
(and charge) unit cells are large at low doping. 

The paper is organized as follows.
The models with orbital degeneracy are introduced in Sec. 
\ref{Sec:2bHm}. The energy of various stripe phases, filled vs. 
half-filled, diagonal vs. vertical, together with the influence of the 
various parameters of the model, are presented in Sec. \ref{Sec:Gse}. 
Their structures are further described in Sec. \ref{Sec:stripes}, 
together with the mechanism which leads to their stability, illustrated
by the double occupancy distribution and by the respective densities of
states. The paper is summarized in Sec. \ref{Sec:Sum}, where we also 
present our main conclusions.

\section{Two-band Hubbard models}
\label{Sec:2bHm}

Even though doped nickelate LSNO is isostructural to its cuprate 
counterpart LSCO, the electronic degrees of freedom in LSNO are more 
involved. In fact, a minimal realistic Hamiltonian for LSNO must 
contain, besides the $x^2-y^2$ orbital states included in the cuprate 
oxide models, also the $3z^2-r^2$ orbital states. Such a model of 
interacting $e_g$ electrons in a 2D ($a,b$) plane may be written as 
follows, 
\begin{equation}
{\cal H}=H_{\rm kin} + H_{\rm int} + H_z,
\label{eq:H_deg}
\end{equation}
with two orbital flavors, $|x\rangle\equiv x^{2}-y^{2}$ and
$|z\rangle\equiv 3z^{2}-r^{2}$, forming a basis in the orbital space. 
The kinetic energy is described by
\begin{equation}
 H_{\rm kin}= \sum_{\langle ij\rangle}\sum_{\alpha\beta\sigma}
           t_{\alpha\beta}
           c^{\dag}_{i\alpha\sigma}c^{}_{j\beta\sigma}, \qquad
\end{equation}
with
\begin{equation}
           t_{\alpha\beta}=-\frac{t}{4}
\begin{pmatrix}
3           & \pm\sqrt{3} \\
\pm\sqrt{3} &  1
\end{pmatrix},
\label{eq:H_kin}
\end{equation}
where $t$ stands for an effective $(dd\sigma)$ hopping matrix element
due to the hybridization with oxygen orbitals on Ni$-$O$-$Ni bonds, and
the off-diagonal hopping $t^{xz}_{ij}$ along $a$ and $b$ axis depends 
on the phase of the $|x\rangle$ orbital along the considered cubic 
direction. The electron-electron interactions contain only on-site 
terms, which we write in the following form,
\begin{eqnarray}
 H_{\rm int}\! &=&
    U\sum_{i}\big( n^{}_{ix\uparrow}n^{}_{ix\downarrow}
                 + n^{}_{iz\uparrow}n^{}_{iz\downarrow}\big)\nonumber \\
    &+&\! \bigl(U-\tfrac{5}{2}J_H\bigr)
      \sum_{i}n^{}_{ix}n^{}_{iz} 
   - 2J_H\sum_{i}\textbf{S}_{ix}\cdot\textbf{S}_{iz}  \nonumber \\
   & +&\! J_H\sum_{i}\bigl( c^{\dag}_{ix\uparrow}c^{\dag}_{ix\downarrow}
                         c^{}_{iz\downarrow}c^{}_{iz\uparrow}
                       + c^{\dag}_{iz\uparrow}c^{\dag}_{iz\downarrow}
                         c^{}_{ix\downarrow}c^{}_{ix\uparrow} \bigr),
\label{eq:H_int}
\end{eqnarray}
where $U$ and $J_H$ stand for the intraorbital Coulomb and Hund's
exchange elements. We also used 
$n_{i\alpha}=\sum_{\sigma}n_{i\alpha\sigma}$ for total electron density 
operators, given by the sum of densities in orbitals $\alpha=x,z$. The 
last term $H_z$ describes the uniform crystal-field splitting between 
$|x\rangle$ and $|z\rangle$ orbitals along the $c$ axis,
\begin{equation}
 H_z = \tfrac{1}{2}E_{z}\sum_{i\sigma} 
(n_{ix\sigma}-n_{iz\sigma}).
\label{eq:H_cf}
\end{equation}
The splitting between the $e_g$ orbitals originates from the tetragonal 
Jahn-Teller distortion of the NiO$_6$ octahedron. In La$_2$NiO$_4$, 
however, the octahedron where the Ni$-$O$-$Ni in-plane (out-of-plane) 
bond lengths are 1.95 (2.26) \AA,\cite{Jor89} respectively, is much 
less distorted as compared to 1.89 and 2.43 \AA\ bonds in La$_2$CuO$_4$,
\cite{Rad94} which reflects the difference in electron filling. In what 
follows we consider only a realistic positive $E_z$ favoring, due to 
elongated octahedra, the $|z\rangle$ electron occupancy over the 
$|x\rangle$ occupancy in doped compounds. 

In order to elucidate the mechanism leading to the stabilization of
the stripes, we also investigate the DDH model. It is obtained using 
the hopping matrix, 
\begin{equation}
t_{\alpha\beta}=-\frac{t}{2}
\begin{pmatrix}
 1  &  0 \\
 0  &  1
\end{pmatrix},
\label{eq:H_kinddh}
\end{equation}
instead of Eq. (\ref{eq:H_kin}). Note, that the total bandwidth is 
$W=6t$ in the $e_g$ model and $W=4t$ in the DDH model (at $E_z=0$), but 
the average diagonal hopping elements are the same --- this ensures 
that electrons are approximately similarly correlated for a given $U$ 
in both cases, while the electron-electron interactions have in the DDH 
model the same form given by Eq.~(\ref{eq:H_int}), provided one 
labels the orbitals as $x$ and $z$.

The model given by Eq. (\ref{eq:H_deg}) is very rich. Even when 
restricting oneself to solutions with only two atoms in the unit cell, 
one obtains an intricate competition between numerous FM and 
AF phases, even on the level of mean field (MF) 
approximation.\cite{Fre05} In particular, we obtained phase diagrams 
showing, in the regime away from half filling, a clear tendency towards 
ferromagnetism for $J_H\simeq U/4$, or at least to the $C$-AF phase for 
$J_H\simeq 0.15U$. 

\begin{figure}[t!]
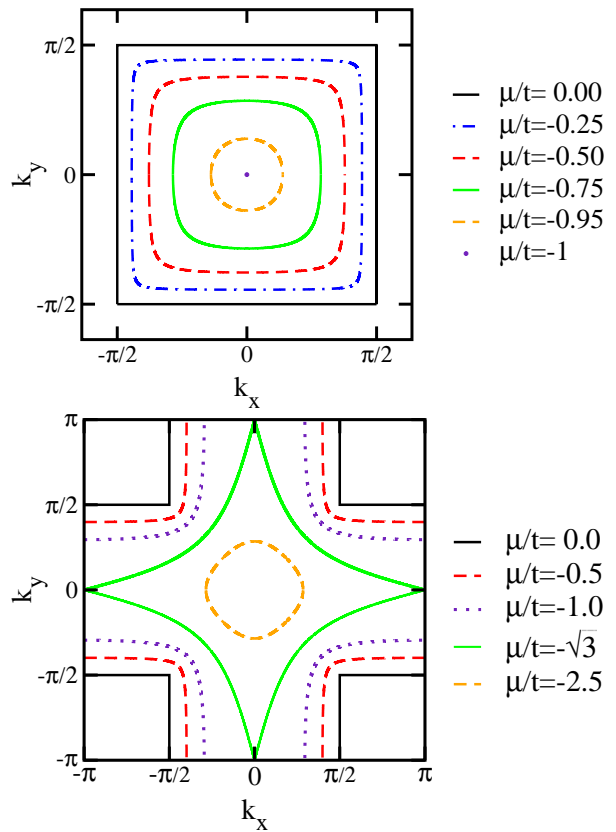

\centerline{\includegraphics*[width=8.0cm]{E_2.eps}}
\centerline{\includegraphics*[width=8.0cm]{E_1.eps}}
\caption{(Color online)
Fermi surfaces of the higher (top) and lower (bottom) bands, as obtained
for different values of the chemical potential $\mu$ in the two-band
tight-binding model for degenerate $e_g$ orbitals ($E_z=0$).
Note a different scale for both Fermi surfaces.
}
\label{fig:FS}
\end{figure}

Some of the richness of the present model (\ref{eq:H_deg}) can be traced 
back to the peculiar features of the electronic dispersion for $e_g$ 
electrons in two dimensions.\cite{Shi98} Unlike for spins, two bands
obtained by diagonalizing the Hamiltonian (\ref{eq:H_deg}) in the 
noninteracting limit are nondegenerate (except for some high symmetry
directions):\cite{noteeg}
\begin{eqnarray}
\varepsilon_{\textbf{k},\nu} &=& -t ( X+Y) \nonumber \\
&+& \nu t \sqrt{X^2 + Y^2 - XY -
  \varepsilon (X+Y) + \varepsilon^2} ,
\label{eq:bands}
\end{eqnarray}
with $\nu=\pm 1$, $X=\cos{k_x}$, $Y=\cos{k_y}$, and 
\begin{equation}
\varepsilon=\frac{E_z}{2t}.
\label{eq:ez}
\end{equation}
For $\varepsilon=0$ the corresponding Fermi surfaces are depicted in 
Fig. \ref{fig:FS}. At half filling one piece of the Fermi surface is 
building a "Swiss cross", while the second one is building a square. 
The filling of the upper band is then by 1/4 electron, while that of
the lower band is by 3/4 electron per spin. The Fermi surface of the 
upper band shrinks rather fast with increasing doping $x$ and below 
$\mu=-t$ ($x=0.93$) one finds that this band is empty and all electrons 
are in the lower band. However, both bands are partly filled in the 
doping regime $x<0.4$ relevant for the nickelates, and they will both 
play a role also when electron interactions are included.

At half filling the edges of the pieces of the Fermi surface (FS) for both 
bands are connected by the nesting vectors: ${\bf Q_1}=(0,\pi)$ and 
${\bf Q_2}=(\pi,0)$. At the same time, the edges of one piece are 
connected to the edges of the other one by the nesting vector 
${\bf Q_3}=(\pi,\pi)$. In addition, 
at half filling both components of the velocities vanish on their 
respective Fermi surfaces. Therefore, several competing instabilities 
are expected once the interaction is turned on. Their relative 
importance can be estimated by computing the spin-spin correlation 
function in the noninteracting limit for the corresponding wave-vectors. 
In the static limit it reads:
\begin{eqnarray}
\chi_S(\textbf{Q})&=& 
\frac{2}{N}\sum_{\textbf{k}}\sum_{\alpha\beta\mu\nu} 
\frac{f_{F}(\varepsilon_{\textbf{k}\mu })
     -f_{F}(\varepsilon_{\textbf{k}+\textbf{Q}\nu})}
     {\varepsilon_{\textbf{k}+\textbf{Q}\nu}
     -\varepsilon_{\textbf{k}\mu}}  \nonumber \\
&\times&U^{}_{\alpha\mu}(\textbf{k})U^{\dag}_{\mu\beta}(\textbf{k})
U^{}_{\beta\nu}(\textbf{k}+\textbf{Q})
U^{\dag}_{\nu\alpha}(\textbf{k}+\textbf{Q})\; .
\end{eqnarray}
Here the matrices $U(\textbf{k})$ diagonalize the hopping matrix
Eq. (\ref{eq:H_kin}), and $\varepsilon_{\textbf{k}\mu}$ are given by
Eq. (\ref{eq:bands}). 

\begin{figure}[b!]
\centerline{\includegraphics*[width=8.0cm]{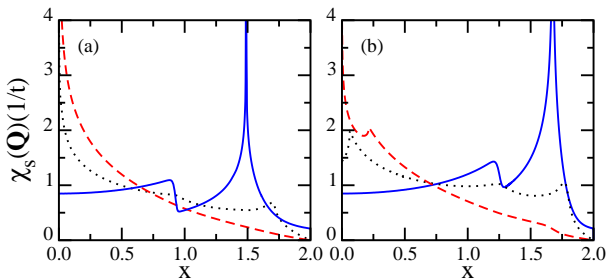}}
\caption
{Static spin-spin correlation functions $\chi_S(\textbf{Q})$ for $e_g$ 
electrons for increasing doping, as obtained for:  
(a) $E_z=0$, and 
(b) $E_z=t$.
Different lines correspond to:
${\bf Q}=(0,0)$ (solid line), ${\bf Q}=(0,\pi)$ (dotted line), and 
${\bf Q}=(\pi,\pi)$ (dashed line).
Van Hove singularity for ${\bf Q}=(0,0)$ is obtained at doping higher
than that realized experimentally in nickelate LSNO oxides:
(a) $x\simeq 1.49$, and (b) $x\simeq 1.68$.
}
\label{fig:bubble}
\end{figure}

Since ferromagnetism is an expected feature of the model 
(\ref{eq:H_deg}), we also evaluate the homogeneous spin-spin correlation 
functions. It shows an intricate interplay of all the three 
instabilities (see Fig.~\ref{fig:bubble}). They 
were all reported, regardless of the approach, see for instance Refs. 
\onlinecite{Hot04} and \onlinecite{Fre05}. One finds that the van Hove 
singularity takes place at $\mu=-\sqrt{3+\varepsilon^2}t$ for the 
crystal-field splitting $\varepsilon$, see Eq. (\ref{eq:ez}). At this 
point the line where the velocity vanishes has a large component on the 
FS, and the singularity is logarithmic. Note that there is no 
nesting vector connecting different pieces of the FS, as can 
be understood by looking at Fig.~\ref{fig:FS}. Here one finds a distinct 
difference between the present $e_g$ model and the DDH model, where 
instead the FS is nested and takes the shape of a square for both bands 
at the van Hove point (for $x=0$). Therefore, nesting and van Hove
singularities in the $e_g$ electron model influence the physical 
properties at different electron densities than in the DDH model. 

In the following the Hamiltonian given by Eq. (\ref{eq:H_deg}) is 
treated in the HF approximation. We introduce explicity four "diagonal" 
expectation values, $\langle n_{i\alpha\sigma}\rangle$, and four 
"off-diagonal" ones, 
$\langle c^{\dag}_{i\alpha\sigma} c^{}_{i\beta\sigma}\rangle$, with 
$\alpha,\beta=x,z$ and $\alpha\neq\beta$, that we determine 
self-consistently in order to get a complete characterization of the 
ground state. Due to symmetry the last four Fock averages vanish for 
a 2D square lattice, and one gets the simpler MF Hamiltonian: 
\begin{equation}
H_{\rm MF} = \sum_{ij,\alpha\beta,\sigma} c^{\dag}_{i\alpha\sigma}
M_{ij,\alpha\beta,\sigma} c^{}_{j\beta\sigma},
\label{eq:H_MF}
\end{equation}
with, 
\begin{equation}
M_{ij,\alpha\beta,\sigma} = t^{\alpha\beta}_{ij} + \delta_{ij}
\delta_{\alpha\beta} M^{}_{i\alpha\sigma},
\end{equation}
and, 
\begin{align}
M^{}_{ix\sigma} &= \tfrac{1}{4}\sum_{i}\big[
                   (3U-5J_H)\langle n_{i}\rangle 
 - \lambda_{\sigma}( U+ J_H)\langle m_{i}\rangle \nonumber \\
&-                 ( U-5J_H)\langle o_{i}\rangle 
 - \lambda_{\sigma}( U- J_H)\langle p_{i}\rangle \big]+ \tfrac{1}{2}E_z,
\nonumber \\
M^{}_{iz\sigma} &= \tfrac{1}{4}\sum_{i}\big[
                   (3U-5J_H)\langle n_{i}\rangle 
 - \lambda_{\sigma}( U+ J_H)\langle m_{i}\rangle \nonumber \\
&+                 ( U-5J_H)\langle o_{i}\rangle 
 + \lambda_{\sigma}( U- J_H)\langle p_{i}\rangle \big]- \tfrac{1}{2}E_z,
\end{align}
where we introduced $\lambda_{\sigma}=\pm 1$ for 
$\sigma=\uparrow(\downarrow)$ spin  and $\lambda_{\alpha}=\pm 1$ for 
$\alpha=x(z)$ orbital, and the operators: 
\begin{align} 
n_i &= \sum_{\alpha\sigma}n_{i\alpha\sigma}, \nonumber \\
m_i & =\sum_{\alpha\sigma}\lambda_{\sigma}n_{i\alpha\sigma}, \nonumber\\
o_i &= \sum_{\alpha\sigma}\lambda_{\alpha}n_{i\alpha\sigma}, \nonumber\\
p_i &= \sum_{\alpha\sigma}\lambda_{\alpha} \lambda_{\sigma}
n_{i\alpha\sigma},
\label{eq:opsdef}
\end{align}
for total charge and magnetization density, and for orbital anisotropy 
in the charge and magnetization distribution, respectively.

\section{Ground state energies}
\label{Sec:Gse}

Site centered and bond centered stripes represent equally relevant 
candidates for the ground state. However, it numerically turns out that 
the energy of SC stripes is substantially larger than the one of the BC
stripes, in agreement with the early study of Yi et al.\cite{Yi98}
(see Table \ref{tab:FEz0} in Sec. \ref{Sec:Sum} as well). We therefore 
concentrate on BC stripes.  
They are made out of pairs of atoms with a FM spin polarization within 
the model Hamiltonian (\ref{eq:H_deg}) in the wide doping regime 
$0.05\leq x\leq 0.4$. In order to obtain unbiased results, we performed 
calculations on large clusters implementing the symmetry of stripe 
phases in reciprocal space. We worked on squared clusters with the 
linear dimension along the $x$ direction chosen as an \emph{even} 
multiplicity of the elementary stripe unit cell dimension. Their sizes 
are listed in Table~\ref{tab:egF}.  
 
\begin{table}[b!]
\caption 
{
Comparison of the ground state free energy $F$ for the VBC (left) and 
DBC (right) stripe phases as found in the $e_g$ model for increasing 
doping $x$. Stripes are separated by $d=3,\dotsc,11$ lattice constants 
in clusters of size $L\times L$. 
Parameters: $U=8t$, $J_H=1.5t$ and $E_z=0$. 
}
\vskip .1cm
\begin{ruledtabular}
\begin{tabular}{cccccccc}
\multicolumn{1}{c}{}              &
\multicolumn{3}{c}{VBC}           &
\multicolumn{1}{c}{}              &
\multicolumn{3}{c}{DBC}            \cr
\multicolumn{1}{c}{$x$}           &
\multicolumn{1}{c}{$d$}           &
\multicolumn{1}{c}{$L\times L$}           &
\multicolumn{1}{c}{$F/t$}         &
\multicolumn{1}{c}{}              &
\multicolumn{1}{c}{$d$}           &
\multicolumn{1}{c}{$L\times L$}           &
\multicolumn{1}{c}{$F/t$}         \cr
\colrule
0.05  &  11  &  88$\times$88 & 2.8811  &&  11  &  88$\times$88 & 2.8794\\
0.06  &  11  &  88$\times$88 & 2.8376  &&  11  &  88$\times$88 & 2.8326\\
0.07  &  11  &  88$\times$88 & 2.7946  &&  11  &  88$\times$88 & 2.7870\\
0.08  &  11  &  88$\times$88 & 2.7519  &&  11  &  88$\times$88 & 2.7430\\
0.09  &  10  &  80$\times$80 & 2.7099  &&  10  &  80$\times$80 & 2.7001\\
0.10  &  10  &  80$\times$80 & 2.6677  &&  9   &  72$\times$72 & 2.6570\\
0.11  &  9   &  72$\times$72 & 2.6258  &&  8   &  64$\times$64 & 2.6140\\
0.12  &  8   &  64$\times$64 & 2.5839  &&  7   &  84$\times$84 & 2.5712\\
0.14  &  7   &  84$\times$84 & 2.5001  &&  6   &  72$\times$72 & 2.4853\\
0.16  &  6   &  72$\times$72 & 2.4162  &&  6   &  72$\times$72 & 2.3997\\
0.18  &  5   &  80$\times$80 & 2.3329  &&  5   &  80$\times$80 & 2.3130\\
0.20  &  5   &  80$\times$80 & 2.2485  &&  4   &  80$\times$80 & 2.2291\\
0.25  &  4   &  80$\times$80 & 2.0394  &&  4   &  80$\times$80 & 2.0155\\
0.30  &  3   &  72$\times$72 & 1.8325  &&  3   &  72$\times$72 & 1.8029\\
0.40  &  3   &  72$\times$72 & 1.4619  &&  3   &  72$\times$72 & 1.4081\\
\end{tabular}
\end{ruledtabular}
\label{tab:egF}
\end{table}

To ensure that the model (\ref{eq:H_deg}) is indeed relevant for the
nickelates, it is necessary to focus on appropriate values of parameters 
$U$, $J_H$, and $E_z$. First of all, x-ray absorption (XAS) measurements 
\cite{Kui91} as well as from XPS and bremsstrahlung-isochromat 
spectroscopy studies \cite{Eis92} of the electronic structure of 
LSNO suggest that LSNO is a CT insulator with nearly the same CT energy 
$\Delta$ as that of NiO. Therefore, the same parameters could be 
accepted as used before in the self-consistent Born calculations,
\cite{Bal00} which reproduced quite well photoemission spectra of 
NiO. Secondly, since our approximation has a tendency to overestimate
the effect of the Coulomb repulsion in LSNO, we used a somewhat smaller 
value of the Coulomb repulsion $U$ than that adequate for NiO, i.e.,  
we set $U=\Delta=5$ eV. Next, as the in-plane Ni$-$O$-$Ni bond length in 
La$_2$NiO$_4$ of 1.95 \AA (see Ref. \onlinecite{Jor89}) is very much the 
same as the shorter bond of 1.89 \AA\ in La$_2$CuO$_4$,\cite{Rad94} we 
set the hopping amplitude $t_{pd}$ between the $p_\sigma$ orbitals and 
$|x\rangle\sim|x^{2}-y^{2}\rangle$ orbitals to be as in LSCO, i.e., 
$t_{pd}=1.47$ eV.\cite{Gra92} This in turn yields an effective in-plane 
Ni$-$Ni hopping $t^{xx}=(t_{pd}^2/\Delta)=0.43$ eV. However, it is more 
convenient to take the effective $(dd\sigma)$ hopping element connecting 
two $|z\rangle\sim|3z^{2}-r^{2}\rangle$ orbitals along the $c$-axis as 
the energy unit $t$, related to $t^{xx}$ via the Slater-Koster relation 
$t=4t^{xx}/3\simeq 0.6$ eV, so that $U\simeq 8t$. 

The value of Hund's exchange between $t_{2g}$ electrons in NiO was
estimated as $J_H'=0.8$ eV.\cite{Zaa90} It is related to $J_H$ for $e_g$ 
electrons through the simple relation,\cite{Ole05}   
\begin{equation}
J_H =J_H' + B,
\label{eq:JH}
\end{equation}
where $B$ stands for a Racah parameter.\cite{Gri71} Taking into account 
that $B\simeq 0.13$ eV for NiO$_2$,\cite{Boc92} one finds $J_H=0.93$ eV, 
i.e., $J_H\simeq 1.5t$. Indeed, it has been shown that $J_H=1$ eV 
reproduces the experimental band gap and the magnetic moment of
La$_2$NiO$_4$.\cite{Mah97} Finally, we set $E_z=t=0.6$ eV as a realistic 
value of the crystal field splitting in the nickelates. On the one hand,
band structure calculations in the local density approximation 
predict the crystal field splitting between $e_g$ orbitals to be 0.5 eV.
\cite{Gra89} On the other hand, XAS spectra reveal a somewhat larger 
splitting of 0.7 eV,\cite{Kui98} a value also deduced from the optical 
spectroscopy.\cite{Jun01}

\begin{figure}[t!]
\centerline{\includegraphics*[width=8.0cm]{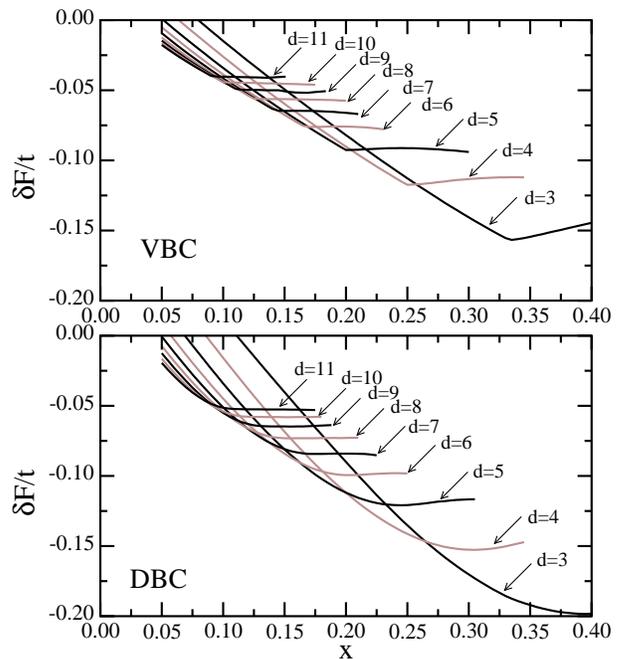}}
\caption
{
Free energy gains $\delta F$ (\ref{eq:df}) of the VBC (top) and DBC 
(bottom) stripe phases with respect to the AF phase (\ref{eq:df}) as 
functions of doping $x$, obtained at temperature $\beta t=100$ for the 
$e_g$ model. Cluster sizes, distances between DWs $d$, and parameters 
are the same as in Table~\ref{tab:egF}.}
\label{fig:F_FAFe_gEz0}
\end{figure}

The performed calculations demonstrate a robust tendency towards stripe 
formation expressed by negative free energy gains of the 
vertical bond-centered (VBC) and diagonal bond-centered (DBC) 
stripe phases with respect to the AF phase,
\begin{equation}
\delta F \equiv F_{S}- F_{AF} \; , 
\label{eq:df}
\end{equation}
for the $e_g$ model (\ref{eq:H_deg}). This energy gain $|\delta F|$ 
increases with doping $x$, as shown in Fig.~\ref{fig:F_FAFe_gEz0}. The
$\delta F$ curves cross each other for decreasing distances between DWs 
$d$ which demonstrates the tendency to the gradual formation of stripe 
phases with smaller unit cells upon increasing doping. Thereby, diagonal 
structures are significantly lower in energy than vertical ones for a 
given fixed doping, but especially in the large doping regime 
$x\sim 0.4$. The robust stability of the DBC stripe phases with respect 
to the VBC ones is illustrated more transparently in Table~\ref{tab:egF}, 
where we compare the ground state free energy $F$ for both structures. 
Note that a similar variation of the optimal distance $d$ between DWs 
suggests the same optimal stripe filling. 

\begin{figure}[t!]
\centerline{\includegraphics*[width=8.0cm]{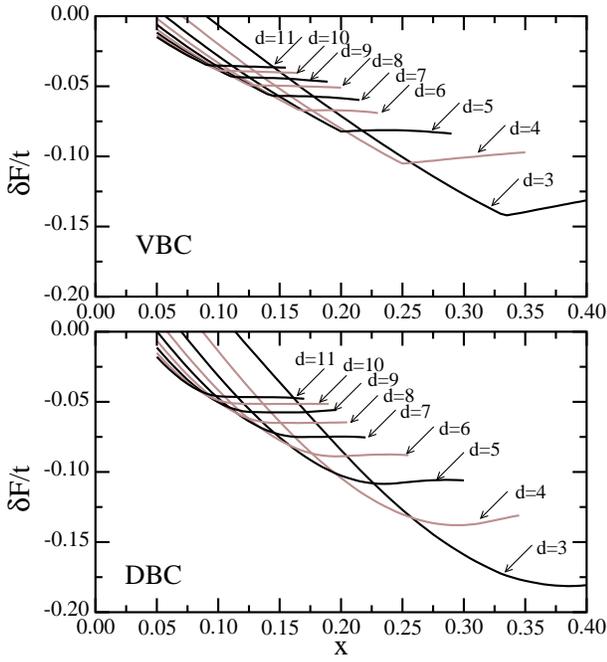}}
\caption
{
Free energy gains $\delta F$ as in Fig.~\ref{fig:F_FAFe_gEz0}, but with 
finite crystal field splitting $E_z=t$. Cluster sizes, distances between 
DWs $d$, and parameters are the same as in Table~\ref{tab:egF}.
}
\label{fig:F_FAFe_gEz1}
\end{figure}

The main effect of the crystal field splitting is to reduce the 
stability of both stripe phases with respect to the uniform AF phase. 
This appears to be rather puzzling at first sight. One finds, however, 
that realistic positive $E_z$ promotes the $|z\rangle$ orbitals with a 
narrow band. On the one hand, the electron distribution induced by 
finite $E_z$ suppresses substantially the AF superexchange energy gain
which becomes $J_{zz}=4t_{zz}^2/U$ rather than $J_{xx}=4t_{xx}^2/U$. 
On the other hand, one would expect that such charge redistribution 
strongly increases the kinetic energy gain due to a wider band 
accessible for holes, propagating especially easily along DWs where the 
AF order is partly suppressed. Nevertheless, this energy gain is easily 
overcompensated by the kinetic energy loss due to hopping perpendicular 
to the stripes. In addition, our studies of the BC stripe phases within 
the single-band Hubbard model have shown that the largest kinetic energy 
gain is released on the bonds connecting pairs of ferromagnetically 
coupled atoms located in the DWs.\cite{Rac06} Note that the 
FM order of the DW spins is substantially stabilized by the off-diagonal 
$t_{xz}$ hopping in the $e_g$ model, yielding low-energy charge 
excitations that lead to the FM superexchange, 
$J_{xz}=4t_{xz}^2/(U-3J_H)$. Altogether, when one orbital is 
sufficiently favored by finite crystal field over the other one, these 
low-energy processes are effectively blocked, explaining the enhanced 
stability of the AF order with respect to the BC solutions 
(\textit{cf}. Figs.~\ref{fig:F_FAFe_gEz0} and \ref{fig:F_FAFe_gEz1}).  
   
A further qualitative point concerns the influence of a finite crystal 
field splitting $E_z$ between the $|x\rangle$ and $|z\rangle$ orbitals 
on the stability of DW structures. As depicted in Fig. 
\ref{fig:F_FAFe_gEz1}, a more realistic value $E_z=t$ seems not to 
promote noticeably any stripe phase over another one and one still 
recovers DBC stripe phases as the ground state. We therefore conclude 
that it is not the crystal field $E_z$ that is responsible for a 
different orientation of DWs in the nickelates from that observed in 
the cuprates. Later on we shall see that finite $E_z$ has also only 
a little visible effect on optimal stripe filling. 

\begin{figure}[t!]
\centerline{\includegraphics*[width=8.0cm]{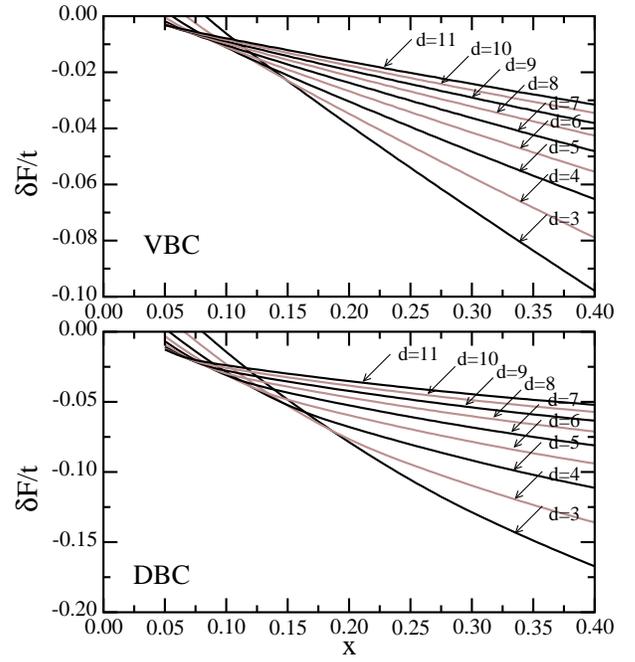}}
\caption{
Free energy gains $\delta F$ of the VBC (top) and DBC (bottom) stripe 
phases with unit cells of length $d$, as a function of doping $x$, 
obtained for the DDH model. 
Parameters: $U=8t$, $J_H=1.5t$, $E_z=0$, and $\beta t=100$. 
}
\label{fig:F_FAFt2_gEz0}
\end{figure}

Let us now shortly return to the experimental characterization of the 
stripe phases in LSNO. Their periodicity at a given doping $x$ is 
described by both charge ${\bf Q}_{c}^{}=\pm\pi(2\epsilon,2\epsilon)$ 
and spin ${\bf Q}_{s}^{}=\pi(1\pm\epsilon,1\pm\epsilon)$ wave vectors 
along the direction diagonal with respect to the Ni$-$O bond. The 
incommensurability $\epsilon$ corresponds to the inverse of the distance 
$d$ between DWs,\cite{Sac95,Tra96,Lee97,Yos00,Kaj01,Lee02,Fre02,Kaj03} 
i.e.,
\begin{equation}
\epsilon=\frac{1}{d}.
\end{equation}
Further, $\epsilon$ first increases continuously with doping $x$ and 
follows the linear relation $\epsilon=x$ in the wide doping range 
$x\leq 1/3$, while next it gradually saturates to the value 
$\epsilon\simeq 0.44$.\cite{Kaj03} Such a behavior of $\epsilon$ 
indicates a fixed hole density of one hole per Ni ion in a DW, and is 
consistent with the HF predictions both in the single and in multiband 
models.\cite{Zaa89,Poi89,Sch89,Kat90,Inu91,Zaa94,Yi98,Miz97} Finally, 
a fixed hole density along the DWs results in the pinning of the 
chemical potential $\mu$ for $x\leq 1/3$, whereas a large ($\simeq 1$ 
eV/hole) downward shift appears in the higher doping regime.\cite{Sat00}

\begin{table}[b!]
\caption 
{Comparison of the ground state free energy $F$ per site for the VBC 
(left) and DBC (right) stripe phases as found in the DDH model for 
increasing doping $x$. Stripes are separated by $d$ lattice constants 
in clusters of size $L\times L$. 
Parameters: $U=8t$, $J_H=1.5t$, and $E_z=0$. 
}
\vskip .1cm
\begin{ruledtabular}
\begin{tabular}{cccccccc}
\multicolumn{1}{c}{}              &
\multicolumn{3}{c}{VBC}           &
\multicolumn{1}{c}{}              &
\multicolumn{3}{c}{DBC}            \cr
\multicolumn{1}{c}{$x$}           &
\multicolumn{1}{c}{$d$}           &
\multicolumn{1}{c}{$L\times L$}           &
\multicolumn{1}{c}{$F/t$}         &
\multicolumn{1}{c}{}              &
\multicolumn{1}{c}{$d$}           &
\multicolumn{1}{c}{$L\times L$}           &
\multicolumn{1}{c}{$F/t$}         \cr
\colrule
0.050  &   11  & 88$\times$88 &  3.1095 && 11  & 88$\times$88 & 3.1000\\
0.055  &   10  & 80$\times$80 &  3.0912 && 11  & 88$\times$88 & 3.0804\\
0.060  &   9   & 72$\times$72 &  3.0729 && 11  & 88$\times$88 & 3.0610\\
0.065  &   8   & 64$\times$64 &  3.0547 && 10  & 80$\times$80 & 3.0417\\
0.070  &   8   & 64$\times$64 &  3.0364 && 9   & 72$\times$72 & 3.0225\\
0.080  &   8   & 64$\times$64 &  3.0001 && 8   & 64$\times$64 & 2.9839\\
0.090  &   7   & 84$\times$84 &  2.9636 && 7   & 84$\times$84 & 2.9454\\
0.100  &   6   & 72$\times$72 &  2.9269 && 7   & 84$\times$84 & 2.9070\\
0.110  &   5   & 80$\times$80 &  2.8902 && 6   & 72$\times$72 & 2.8683\\
0.120  &   4   & 80$\times$80 &  2.8538 && 5   & 80$\times$80 & 2.8301\\
0.140  &   4   & 80$\times$80 &  2.7807 && 5   & 80$\times$80 & 2.7530\\
0.160  &   3   & 72$\times$72 &  2.7077 && 4   & 80$\times$80 & 2.6760\\
0.180  &   3   & 72$\times$72 &  2.6347 && 4   & 80$\times$80 & 2.5996\\
0.200  &   3   & 72$\times$72 &  2.5626 && 3   & 72$\times$72 & 2.5227\\
0.300  &   3   & 72$\times$72 &  2.2129 && 3   & 72$\times$72 & 2.1531\\
0.400  &   3   & 72$\times$72 &  1.8809 && 3   & 72$\times$72 & 1.8116\\
\end{tabular}
\end{ruledtabular}
\label{tab:DDHF}
\end{table}

It is interesting to establish whether the above results concerning the 
stripe phases in the ground state and their variation under increasing 
doping appear solely in the realistic $e_g$ model, or they are generic 
and remain also a common feature of the DDH model with only diagonal 
hopping elements given by Eq. (\ref{eq:H_kinddh}). To allow a meaningful 
comparison, we set the same values of microscopic parameters as we those 
chosen for the $e_g$ model, i.e., $U=8t$, $J_H=1.5t$, and we consider 
only the $E_z=0$ case imposed by symmetry between the equivalent 
orbitals. It restricts the solutions of the DDH model to 
$\langle o\rangle=\langle p\rangle=0$ (see below). 

One finds that the free energy gains $\delta F$ of the VBC and DBC 
stripe phases with respect to the AF phase for the DDH model (Fig. 
\ref{fig:F_FAFt2_gEz0}) are qualitatively similar to those of the $e_g$ 
band model. First of all, for a fixed doping $x$ diagonal stripe 
structures are again significantly lower in energy than vertical ones 
(\textit{cf}. also Table~\ref{tab:DDHF}). Secondly, also in this case, 
we recover a gradual crossover towards stripe phases with smaller unit 
cells upon increasing doping. Note, however, that in contrast to the 
predictions made within the $e_g$ model, structures with vertical 
(diagonal) DWs separated by a distance $d\geq 5$ are the lowest energy 
solutions only in a narrow doping regime $x\lesssim 0.12$ ($x\lesssim 
0.15$), respectively. 

\begin{figure}[t!]
\centerline{\includegraphics*[width=8.0cm]{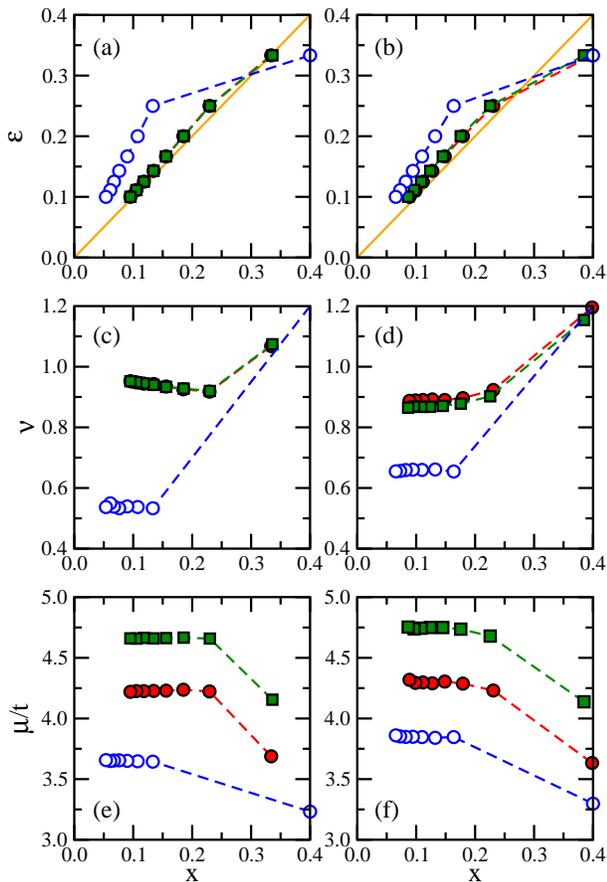}}
\caption{(Color online)
Doping dependence of stripe phases for the VBC (left) and DBC (right)
ground states: 
(a) and (b) magnetic incommensurability $\epsilon$; 
(c) and (d) stripe filling $\nu$; 
(e) and (f) chemical potential $\mu$ (in the units of $t$),  
as deduced from the data of Figs.~\ref{fig:F_FAFe_gEz0}, 
\ref{fig:F_FAFe_gEz1} and \ref{fig:F_FAFt2_gEz0}.
Filled (open) symbols for $e_g$ (DDH) model; 
circles and squares for $E_z=0$ and $E_z=t$. 
Solid lines in panels (a) and (b) show the experimental behavior of 
$\epsilon$ in LSNO given in Ref.~\onlinecite{Kaj03}. 
}
\label{fig:enme_g}
\end{figure}

For a complete characterization of stripe phases we define
\begin{equation}
\nu=\frac{N_h}{L N_{\rm DW}},
\end{equation}
where $N_h$ is the number of doped holes and $N_{\rm DW}$ is the number 
of domain walls. Our findings concerning the properties of 
BC stripe phases are summarized in Fig.~\ref{fig:enme_g} showing the 
doping dependence of the incommensurability $\epsilon$, the stripe 
filling $\nu$, and the chemical potential $\mu$ for both the VBC (left) 
and DBC (right) stripe phases. These quantities for the respective 
ground states are deduced from Figs. \ref{fig:F_FAFe_gEz0}, 
\ref{fig:F_FAFe_gEz1} and \ref{fig:F_FAFt2_gEz0}. In each case the data 
points correspond to the middle of the stability region of the stripe 
phase of a given type. The only exception are the $d=3$ phases --- for 
them $\epsilon$, $\nu$, and $\mu$, are plotted for the actual minimum 
of the free energy $F$. 

In agreement with the experimental data for LSNO, indicated here by a 
solid line in panels (a) and (b) of Fig. \ref{fig:enme_g}, one 
observes that $\epsilon$ follows the law $\epsilon\simeq x$ up to 
$x\simeq 0.2$ and then it tends to saturate to the highest possible 
value for the BC stripe phase, i.e., $\epsilon=1/3$, as there are no 
BC stripe phases in which DWs are separated by $d=2$ lattice spacings. 
Further, in the regime where $\epsilon$ follows linearly the doping 
level $x$, decreasing stripe periodicity allows the system to maintain 
nearly fixed filling $\nu$ [\textit{cf}. Figs. \ref{fig:enme_g}(c) and 
\ref{fig:enme_g}(d)], pinning simultaneously the chemical potential 
$\mu$, as shown in Figs. \ref{fig:enme_g}(e) and \ref{fig:enme_g}(f). 
Remarkably, the ground state of both the VBC and DBC stripe phases is 
characterized by the optimal filling 0.9 hole/Ni, very close indeed to 
the experimental value one hole/Ni ion, and the optimal filling remains 
almost unaltered in the model with a finite crystal field $E_z=t$. 

Regarding the chemical potential shift $\Delta\mu$ in the doping regime 
$x>0.2$, one finds that it exceeds the experimental value $\sim -1.0$ 
eV/hole nearly by a factor of 2. Indeed, assuming the effective hopping 
$t=0.6$ eV, one obtains $\Delta\mu\simeq -2.2$ ($-2.1$) eV/hole for  
$E_z=0$ ($E_z=t$), respectively. Therefore, we conclude that the present
effective model can only explain qualitative trends and one needs to 
carry out calculations within more realistic multiband models with 
oxygen orbitals included explicitly in order to obtain quantitatively 
the experimental data.

Let us verify now whether the established results concerning the doping 
dependence of $\epsilon$ and $\nu$ are also obtained within the DDH 
model. As the structures with vertical (diagonal) DWs separated by a 
distance $d\geq 5$ appear in the DDH model only in a narrow doping 
regime $x\lesssim 0.12$ ($x\lesssim 0.15$), respectively, one observes 
here a fast variation of the optimal distance $d$ which should result 
in a small optimal stripe filling. Indeed, as depicted in Figs. 
\ref{fig:enme_g}(a) and \ref{fig:enme_g}(b), the low-doping part of 
$\epsilon$, being linear in $x$, has a larger slope which exceeds the 
experimental value in LSNO roughly by a factor of 2 (1.5) in the case 
of the VBC (DBC) stripe phase, respectively. Consequently, the optimal 
stripe filling in the former case is substantially reduced down to 
$\nu\simeq 0.55$ and in the latter case --- down to $\nu\simeq 0.65$ 
[see Figs.~\ref{fig:enme_g}(c) and \ref{fig:enme_g}(d)]. Finally, one 
finds that the chemical potential starts to decrease rapidly at 
$x\simeq 0.14$ in the VBC stripe phase [Fig.~\ref{fig:enme_g}(e)] and 
at slightly larger doping $x\simeq 0.16$ in the DBC phase
[Fig.~\ref{fig:enme_g}(f)].

\section{Mechanism of stripe formation}
\label{Sec:stripes} 

\subsection{Charge and magnetization distributions in the stripe phases}
\label{Sec:SoS} 

\begin{figure}[t!]
\centerline{\includegraphics*[width=8.0cm]{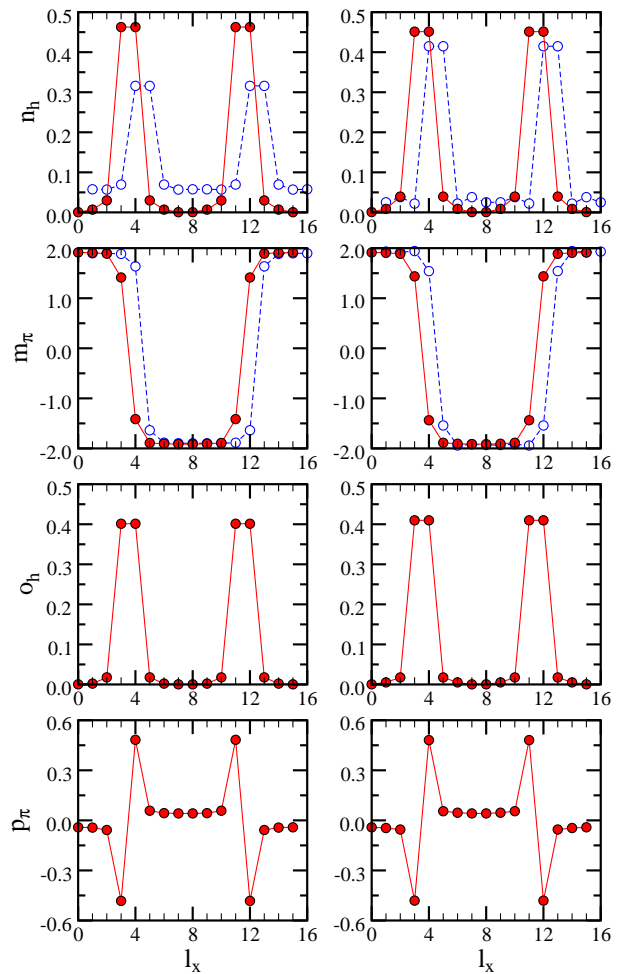}}
\caption{(Color online)
Charge and magnetization distribution in the filled VBC (left) and DBC 
(right) stripe phase found in either the $e_g$ model (filled circles), 
or in the DDH model (open circles) on a 
$64\times 64$ cluster at doping $x=1/8$:
local hole densities $n_h(l_x)$ (top row); 
modulated magnetization densities $m_{\pi}(l_x)$ (second row);
local hole $o_{h}(l_x)$ (third row); 
and local modulated magnetic $p_{\pi}(l_x)$ orbital polarization 
(bottom row). For more clarity, the data points for the DDH model are 
shifted by one lattice constant from the origin of the coordinate 
system.
Parameters: $U=8t$, $J_H=1.5t$, $E_z=0$, and $\beta t=100$.
}
\label{fig:FBC}
\end{figure}

In order to find out the reason of such a vast discrepancy between the 
predictions made in the $e_g$ and in the DDH model, let us now 
investigate closer the properties of idealized filled and half-filled 
DWs: half-filled vertical bond-centered (HVBC) and half-filled diagonal 
bond-centered (HDBC). 
A complete characterization of the charge and magnetization
distributions for a system with orbital degeneracy is given by:
the local hole density $x(l_x)$, 
local modulated magnetization density 
$m_{\pi}(l_x)$, local hole orbital polarization $o_{h}(l_x)$, and 
local modulated magnetic orbital polarization $p_{\pi}(l_x)$, defined
using the operators given in Eqs. (\ref{eq:opsdef}) as follows:
\begin{align}
    n_h(l_x) &= \sum_{\alpha}n_{h\alpha}(l_x),            \nonumber \\               
m_{\pi}(l_x) &=(-1)^{l_x}
\sum_{\alpha\sigma}\lambda_{\sigma} n_{\alpha\sigma}(l_x),\nonumber \\
  o_{h}(l_x) &= \sum_{\alpha}\lambda_{\alpha}n_{h\alpha}(l_x),\nonumber \\
p_{\pi}(l_x) &=(-1)^{l_x}\sum_{\alpha\sigma}\lambda_{\alpha}
                \lambda_{\sigma}  n_{\alpha\sigma}(l_x).
\label{eq:nmopav}
\end{align}
Here $n_{\alpha\sigma}(l_x)$ is the local orbital charge density for
spin $\sigma$, 
whereas $n_{h\alpha}(l_x)$ denotes the local orbital hole density,
\begin{align}
 n_{h\alpha}(l_x)= 1 - \sum_{\sigma}n_{\alpha\sigma}(l_x).
\label{eq:nhal}
\end{align}

\begin{figure}[t!]
\centerline{\includegraphics*[width=8.0cm]{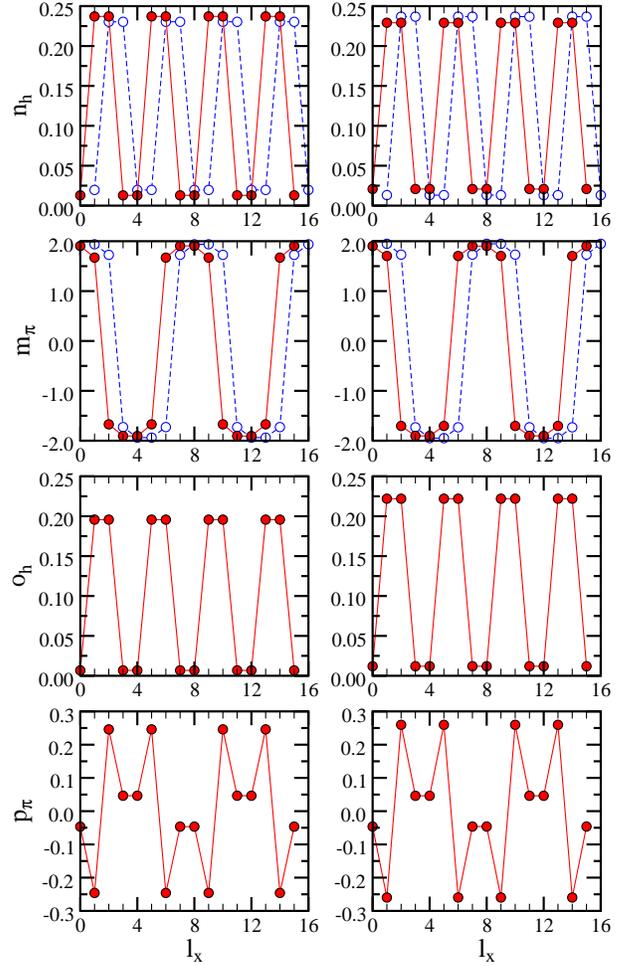}}
\caption{(Color online)
Charge and magnetization distributions in the half-filled HVBC (left) 
and HDBC (right) stripe phases found in 
either the $e_g$ (filled circles) model, or in the DDH model (open 
circles) on a $64\times 64$ cluster at doping $x=1/8$.
The meaning of the different panels and shift of the DDH data
as in Fig. \ref{fig:FBC}.
Parameters: $U=8t$, $J_H=1.5t$ $E_z=0$, and $\beta t=100$.
}
\label{fig:HFBC}
\end{figure}

To better appreciate the differences between the DDH and the $e_g$ 
model, we compare in Fig.~\ref{fig:FBC}, the local hole $n_h(l_x)$  
and the modulated magnetization densities $m_{\pi}(l_x)$ of the filled
VBC and DBC stripe phases found at temperature $\beta t=100$ in either 
model on a $64\times 64$ cluster at doping $x=1/8$ for the standard 
parameters, {\it i.e.}, $U=8t$, $J_H=1.5t$, and $E_z=0$. For 
completeness we also show the local hole $o_{h}(l_x)$ and the local 
modulated magnetic $p_{\pi}(l_x)$ orbital polarization. These quantities 
are finite only in the $e_g$ model, while $o_{h}(l_x)=p_{\pi}(l_x)=0$ by 
symmetry in the DDH model. Since $o_h$ is positive the holes are located 
in $|x\rangle$ orbitals, and the polarization of the two sites on the 
DWs is FM (alternation of signs in $p_{\pi}$ at the DWs). 

The observed differences, especially pronounced at DW atoms, directly 
follow from the fact that, a different effect helping to reduce double 
occupancy at those sites is effective in each model. Namely, in the 
$e_g$ hopping model (\ref{eq:H_kin}) one finds large positive 
$o_{h}(l_x)$ at DWs, which means that it is energetically advantageous 
to optimize the kinetic energy of holes by putting them into the 
$|x\rangle$ orbitals where the hopping element $t_{xx}$ is larger and 
gives a wide band. On the one hand, in the HF the only way to optimize 
the on-site energy is to develop a strong spin polarization which, 
in turn, would noticeably reduce the kinetic energy gain. 
On the other hand, such disadvantageous suppression 
can be avoided by a strong reduction of the electron density. This 
explains why the  hole density $n_h(l_x)$ along DWs in the $e_g$ model 
is larger, as compared to the corresponding value found in the DDH case. 
Indeed, in the latter case both bands are equivalent, resulting in 
$o_{h}(l_x)=p_{\pi}(l_x)=0$. Hence this model yields a more localized 
stripe phase with a larger magnetization $|m_{\pi}(l_x)|$ at DWs than 
the one obtained in the $e_g$ model (\textit{cf}. Fig.~\ref{fig:FBC}).

For completeness, in Fig.~\ref{fig:HFBC} we compare the hole density,
$n_h(l_x)$, and the spin density, $m_{\pi}(l_x)$, profiles of the  
half-filled stripe phases. The unit cells are smaller by a factor of two
and the amplitude of the charge density wave is correspondingly reduced. 
Even though the overall shape of the DWs looks very much the same in 
both models, a larger magnetization $|m_{\pi}(l_x)|$ at DWs is found 
again within the DDH Hamiltonian (\ref{eq:H_kinddh}).

\subsection{Double occupancy distribution}
\label{Sec:double} 

\begin{figure}[t!]
\centerline{\includegraphics*[width=8.0cm]{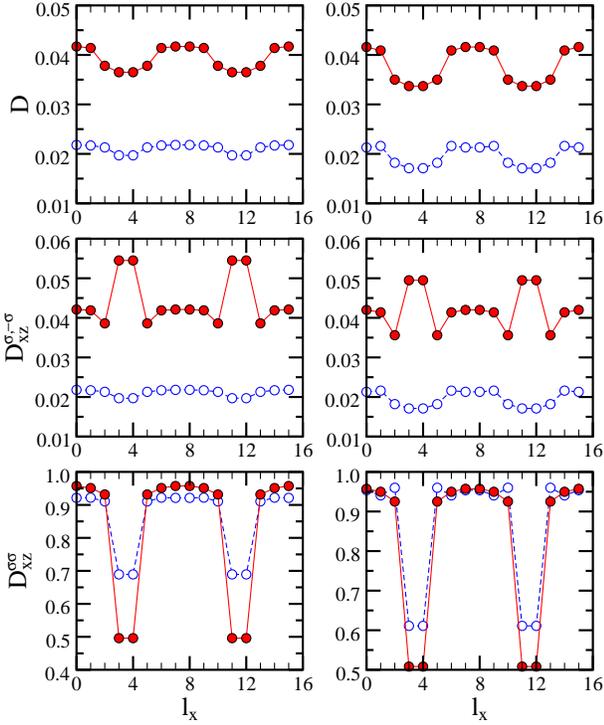}}
\caption{(Color online)
Double occupancy distribution in the filled VBC (left) and DBC (right) 
stripe phase shown in Fig.~\ref{fig:FBC}:
intraorbital double occupancy $D(l_x)$ (top row), and 
two local interorbital $D_{xz}^{\sigma\bar{\sigma}}(l_x)$ (middle) and  
$D_{xz}^{\sigma\sigma}(l_x)$ (bottom) double occupancies. Filled (open) 
circles denote the results found in the $e_g$ (DDH) model, respectively. 
Parameters and filled/empty circles as in Fig. \ref{fig:FBC}.
}
\label{fig:FBCDO}
\end{figure}

Furthermore, important information about the nature of stripe phases is 
provided by the averaged intraorbital double occupancy at site $l_x$,
\begin{equation}
D(l_x) =
\sum_{\alpha} n_{\alpha\uparrow}(l_x)n_{\alpha\downarrow}(l_x),
\label{eq:D0}
\end{equation}
as well as by two local interorbital double occupancies,
\begin{align}
D_{xz}^{\sigma\bar{\sigma}}(l_x) & =
\sum_{\sigma} n_{x\sigma}(l_x)n_{z\bar{\sigma}}(l_x),
\label{eq:DS} \\
D_{xz}^{\sigma\sigma}(l_x) & =
\sum_{\sigma} n_{x\sigma}(l_x)n_{z\sigma}(l_x),
\label{eq:DT}
\end{align}
where $\bar{\sigma}=-\sigma$. In a system with isotropic charge 
distribution the double occupancies are site independent, while in
stripe phases they vary with a characteristic periodicity following 
from the size of the stripe unit cell. From the structure of the 
Coulomb interaction Eq.~(\ref{eq:H_int}) one expects
$D<D_{xz}^{\sigma\bar{\sigma}}<D_{xz}^{\sigma\sigma}$ for a fixed hole 
density. In a stripe phase these double occupancies are locally
suppressed by the magnetic and orbital polarizations in a way that 
depends both on the shape of the stripe, and on the form of the 
one-electron Hamiltonian. Let us first discuss the average double 
occupancies (\ref{eq:D0})-(\ref{eq:DT}) of the stripe phases depicted 
in Fig.~\ref{fig:FBC}. Remarkably, in the DDH model (open circles in 
Fig.~\ref{fig:FBCDO}) the on-site energy is predominantly optimized by 
the reduction of the high-energy configurations with opposite spins 
which lead to intraorbital $D(l_x)$ and interorbital 
$D_{xz}^{\sigma\bar{\sigma}}(l_x)$ double occupancies, so that the 
system might create a large number of DWs in the unit cell even in the 
low doping regime, consequently reducing the optimal stripe filling.  
Such a strong reduction of the double occupancies for opposite spins 
follows from the diagonal hopping which gives a weaker electronic 
mixing between the sites of opposite spin due to a lower kinetic energy.
\cite{Fei05}

\begin{figure}[t!]
\centerline{\includegraphics*[width=8.0cm]{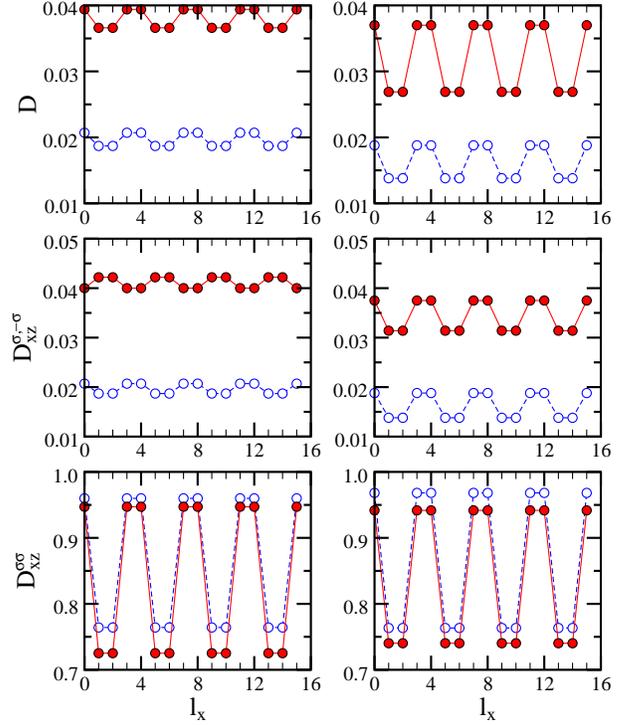}}
\caption{(Color online)
Double occupancy distribution in the half-filled stripe phases
shown in Fig.~\ref{fig:HFBC}: HVBC (left) and HDBC (right).
Different double occupancies as in Fig. \ref{fig:FBCDO}. 
Parameters and filled/empty circles as in Fig.~\ref{fig:HFBC}.
}
\label{fig:HFBCDO}
\end{figure}

In contrast, in the $e_g$ model (filled circles in Fig. 
\ref{fig:FBCDO}), the Coulomb energy is mainly optimized by the 
reduction of the low-energy interorbital $D_{xz}^{\sigma\sigma}(l_x)$ 
double occupancy resulting in a smaller magnetization $|m_{\pi}(l_x)|$ 
at DWs as compared to the one found in the DDH model. However, reduced  
$|m_{\pi}(l_x)|$ at those sites allows the system to better optimize  
the kinetic energy gain which then overcompensates a large on-site 
energy only when the optimal filling is close to one hole per Ni site,
meaning that for a given doping level, the DWs should be separated by a 
larger distance as compared to predictions made in the DDH model. On the 
other hand, the robust stability of the DBC stripe phases with respect 
the VBC ones can be understood as following from a stronger reduction of 
all double occupancies (\ref{eq:D0})-(\ref{eq:DT}) in the former phase.
In fact, the doubly occupied configurations cannot be then excited 
between the nearest neighbor DW sites, leading to a lower total energy.   

In case of half-filled stripes (Fig. \ref{fig:HFBCDO}) the double 
occupancies for opposite spins ($D(l_x)$ and 
$D_{xz}^{\sigma\bar{\sigma}}(l_x)$) are only slightly lower than for 
filled stripes, but the interorbital double occupancy for equal spins 
($D_{xz}^{\sigma\sigma}(l_x)$) is significantly higher, reflecting the 
higher electron filling at DW atoms in these phases. Due to the finite 
Hund's exchange considered here, $J_H=0.15t$, the high-spin states are
promoted. Altogether one finds similar generic features seen above for 
the filled stripes (Fig.~\ref{fig:FBCDO}): 
 (i) both double occupancies for the opposite spins are stronger reduced 
     in the DDH model, and 
(ii) the double occupancy by the same spin $D_{xz}^{\sigma\sigma}$ at 
     the DWs is stronger suppressed in the $e_g$ model, resulting from 
     a smaller magnetization at these sites than in the DDH model.
This second effects seems to play a role in stabilizing the HDBC
stripe phase in the DDH model rather than the DBC phase, found for the
realistic $e_g$ model.     

\subsection{Densities of states}
\label{Sec:dos} 

\begin{figure}[t!]
\centerline{\includegraphics*[width=8.0cm]{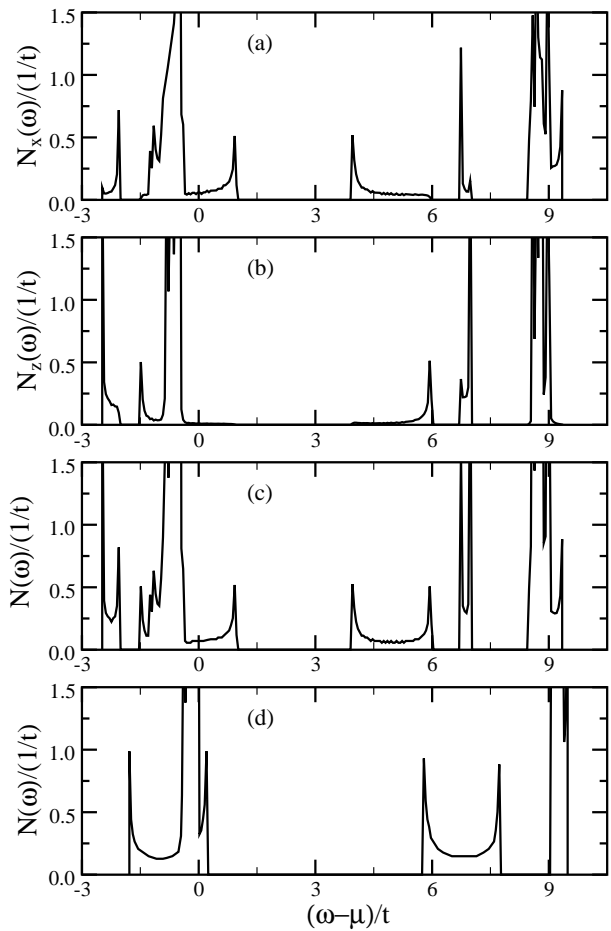}}
\caption
{
Densities of states for the DBC stripe phase in the $e_g$ model at 
$x=1/8$ doping: 
(a) partial density of states $N_x(\omega)$ for $|x\rangle$ orbital, 
(b) partial density of states $N_z(\omega)$ for $|z\rangle$ orbital, 
(c) total density of states $N(\omega)$.
Panel (d) shows for comparison the total density of states $N(\omega)$
as found in the DBC stripe phase within in the DDH model. 
Parameters as in Fig.~\ref{fig:FBC}.
}
\label{fig:DOSF}
\end{figure}

\begin{figure}[t!]
\centerline{\includegraphics*[width=8.0cm]{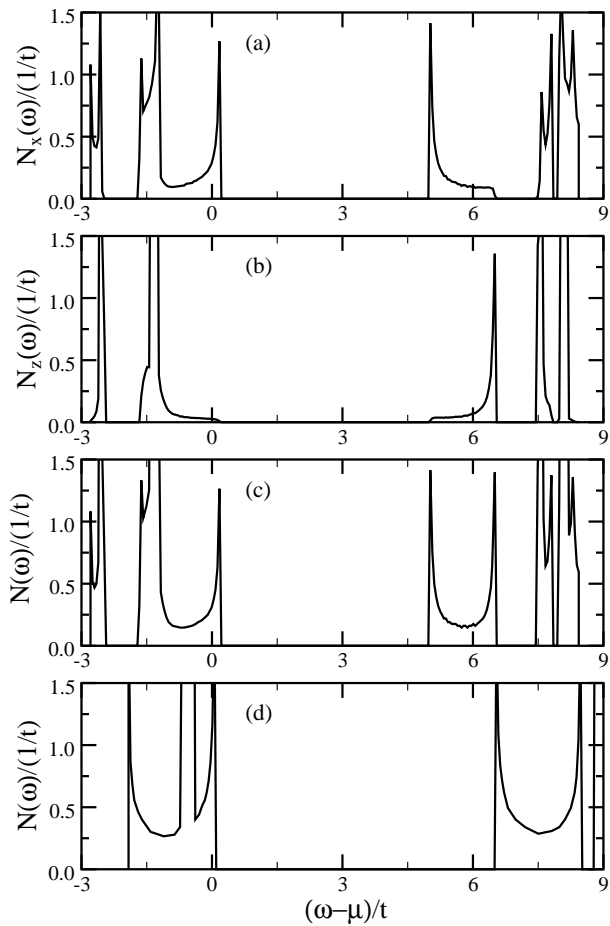}}
\caption
{
Densities of states for the HDBC stripe phase in the $e_g$ model at 
$x=1/8$ doping, panels (a)-(c) as in Fig. \ref{fig:DOSF}.
Panel (d) shows for comparison the total density of states $N(\omega)$
as found in the HDBC stripe phase within in the DDH model. 
Parameters as in Fig.~\ref{fig:HFBC}.
}
\label{fig:DOSHF}
\end{figure}

Important differences between the two models strongly influence the 
density of states (DOS). Here we first determine the DOS projected on
the $\alpha$ orbital as:
\begin{equation}
N_{\alpha}(\omega) = \frac{1}{N}\sum_{\textbf{k}}\sum_{i\sigma}
|\Psi^{}_{i\alpha\sigma}(\textbf{k})|^2
\delta(\omega - \varepsilon_{\textbf{k}\sigma}),
\label{eq:DOSal}
\end{equation}
where $\Psi^{}_{i\alpha\sigma}(\textbf{k})$ are the eigenvectors of the
MF Hamiltonian Eq.~(\ref{eq:H_MF}). It is calculated using a histogram 
of the corresponding eigenvalues. It is apparent from Figs. 
\ref{fig:DOSF} and \ref{fig:DOSHF} that the spectra consist of several 
subbands. However, two Hubbard subbands are well visible both for 
filled and half-filled diagonal BC stripes: lower Hubbard band (LHB) 
and upper Hubbard band (UHB). The Hubbard subbands arise primarily due 
to the AF polarization and the gap between them is proportional to the 
magnetization. 

Panels (a) and (b) in Figs. 
\ref{fig:DOSF} and \ref{fig:DOSHF} show the DOS $N_{\alpha}(\omega)$ 
(\ref{eq:DOSal}), projected on the orbital $|x\rangle$ and $|z\rangle$, 
respectively. Here, one finds that the mid-gap bands have mainly   
$|x\rangle$ character close to the gap between them --- the $|x\rangle$ 
orbitals optimize the kinetic energy of holes doped into them, whereas 
the vast majority of the more localized $|z\rangle$ states of DW atoms 
belongs to the energy regimes within the LHB and close to the UHB. 

The total DOS for the DBC stripes in the $e_g$ model, depicted in 
Fig.~\ref{fig:DOSF}(c), reveals mid-gap states which demonstrate the
one-dimensional character of the transport in the stripe phases. 
Note that due to the AF spin modulation along the BC DWs themselves, 
one finds as well a distinct gap between two mid-gap bands lying within 
the Mott-Hubbard gap. This gap for the filled DBC stipes is lower 
in the $e_g$ model [Fig. \ref{fig:DOSF}(c)] than in the DDH model [Fig. 
\ref{fig:DOSF}(d)], as the holes are more delocalized in the former 
case. This holds as well for the HDBC stripes, see Figs. 
\ref{fig:DOSHF}(c) and \ref{fig:DOSHF}(d). We have shown before that
the magnetization on the DW sites is larger for the HDBC stripes than
for the DBC ones, and therefore the gaps between the mid-gap states are 
also larger in Fig. \ref{fig:DOSHF} than in Fig. \ref{fig:DOSF}.

Remarkably, both for the filled and half-filled diagonal BC stripe 
phases of Figs. \ref{fig:DOSF} and \ref{fig:DOSHF} the lower mid-gap 
states are to some extent
localized above the LHB in the $e_g$ model, whereas they are clearly 
almost merged with the LHB in the DDH model. This behavior follows 
from a smaller on-wall magnetization $m_{\pi}(l_x)$ in the $e_g$ model.   
Therefore, in the DDH case only a few holes occupy the low-lying 
mid-gap bands, explaining the reason of substantial suppression of the 
optimal stripe filling in this case. Instead, filled stripes are favored 
in the $e_g$ model --- they have a larger fraction of holes in the lower 
mid-gap states. Note,
however, that also in this case the stripes are metallic, with some
spectral weight (mainly in $|x\rangle$ orbitals) above $\omega=\mu$.

\section{Discussion and summary}
\label{Sec:Sum}

We summarize our key results concerning the stability of stripe phases
in the nickelates. We investigated this problem using two Hubbard models 
--- the realistic model for $e_g$ electrons, and the DDH model which 
frequently serves as the simplest approach to the orbital degeneracy.
\cite{Kle98} 
It is quite encouraging that the DDH model gives already diagonal stripe 
phases as the most stable structures, in contrast to the nondegenerate 
Hubbard model where the vertical (horizontal) stripes are found in the 
HF approximation. However, one finds half-filled BC stripes in the DDH 
model, and it is only in the realistic $e_g$ model for doped nickelates 
that one observes a generic tendency to promoting filled stripes over 
the half-filled ones. As an example of this behavior, we give the free
energies of filled and half-filled stripe phases found within both 
($e_g$ and DDH) models for $x=1/8$ doping in Table~\ref{tab:FEz0}. 
First of all, one finds that the energies of BC phases are considerably 
lower for the realistic model of $e_g$ band due to the higher kinetic 
energy gains which result from the off-diagonal hopping.\cite{Fei05}
Most importantly, for the parameters relevant to LSNO used for the data
of Table~\ref{tab:FEz0}, one finds that the DBC phase has the lowest 
energy, {\it i.e.\/}, indeed experimentally observed \emph{filled} 
diagonal stripes. In contrast, the SC stripe phases with unpolarized DW
sites and large double occupancies have much higher energies, and the 
best of them is not diagonal, but vertical (horizontal) one. We also 
note that even though the DDH model with two equivalent orbitals 
clearly favors diagonal DWs, it stabilizes instead of filled the 
\emph{half-filled} diagonal phases. 

\begin{table}[b!]
\caption{
Comparison of the free energy $F$ (in the units of $t$) for the BC and SC 
stripe phases in the $e_g$ model, and for BC stripe phases in the DDH 
model, obtained on $64\times 64$ clusters at $x=1/8$ hole doping. The 
free energy of the most stable phase (for BC or SC phases) is given 
in bold characters in each case.
Parameters: $U=8t$, $J_H=1.5t$, $E_z=0$, and $\beta t=100$. 
}
\vskip .1cm
\begin{ruledtabular}
\begin{tabular}{cccccc}
 &  & \multicolumn{2}{c}{diagonal stripes} & 
      \multicolumn{2}{c}{vertical stripes} \cr
 model & phase & filled & half-filled & filled & half-filled \cr
\colrule
$e_g$ & BC &{\bf 2.5508} &     2.5759  &     2.5629  & 2.5756 \cr
      & SC &     2.7815  &     3.2615  &{\bf 2.7012} & 3.0729 \cr
 DDH  & BC &     2.8172  &{\bf 2.8135} &     2.8395  & 2.8354 \cr
\end{tabular}
\end{ruledtabular}
\label{tab:FEz0}
\end{table}

As in the cuprates, the coexisting charge and magnetic order in 
diagonal stripe phases in the nickelates is a result of the 
compromise between the kinetic and magnetic energies --- the magnetic
energy is gained in the AF domains, and the kinetic energy is gained 
mainly along the DWs. The BC stripes are favored as then the magnetic
energy can be gained (Table \ref{tab:FEz0}) not only in the AF 
domains, but also on the DW magnetic sites. Finally, when the DWs
are filled, more kinetic energy is gained in the $e_g$ model while 
then the off-diagonal hopping is allowed. Altogether, this mechanism
shows little sensitivity to the small crystal field splitting. For 
moderate values of the latter it tends to eliminate stripe phases 
entirely rather than promoting one stripe phase over another one.  

In spite of the remarkable success of the present study which gave 
stable diagonal stripe phases, one has to admit that the predicted 
electronic properties show some difference to the experimental ones. 
In fact, our systematic studies of stripe phases completed within the 
relevant model for $e_g$ orbitals, where several phases separated by 
different lattice spacing varying from $d=3$ to $d=11$ were considered, 
have revealed that the optimal stripe filling in the true ground state 
is slightly less ($0.86\leq n_h\leq 0.89$ hole/Ni depending on the 
crystal field splitting) than the experimental value of one  hole per 
Ni ion. This concerns the entire low doping regime, $x\leq 0.3$, where 
$\epsilon\simeq x$. There may be a few reasons of this discrepancy.
First of all, one has to realize that in a multiband model with oxygen 
orbitals included explicitly the holes would be doped primarily to 
oxygen orbitals, screening the local moments at Ni sites,\cite{Zaa90} 
in analogy to Zhang-Rice singlets in the cuprates.\cite{Zha88} An 
insulating ground state could then result from oxygen distortions by 
the Peierls mechanism. Thus, we argue that the present results could be 
further improved within a realistic model including not only two $e_g$ 
orbitals with different hopping elements, but also orbital polarization 
at oxygen sites.
Second, it may be expected that the stripe filling will be somewhat 
changed due to the electron correlation effects beyond the HF 
approximation used here. We believe that the present work provides a
good starting point for future studies of the correlation effects.

In summary we have analyzed the stripe formation in the doped 
nickelates in the realistic model with degeneracy of $e_g$ orbitals,  
using large clusters. The results obtained with this model were
compared with the widely used doubly degenerate Hubbard model. 
Even though in both models the distance between the DW is inversely 
proportional to the doping, the most stable diagonal stripes are found 
to differ markedly from one model to the other one. In the $e_g$ model, 
the stripes are filled and nearly insulating, as observed 
experimentally in a series of layered nickelates. In contrast, 
in the DDH model they are half-filled and metallic. These latter 
stripe phases are reminiscent of the stripes observed experimentally
in largely doped cuprates. These differences have their roots in the 
different structure of intersite hopping terms. As the DDH model is 
closer to $t_{2g}$ than to $e_g$ hopping matrix elements, one might 
expect that doped insulators with $t_{2g}$ orbital degrees of freedom,
as in vanadates or in ruthenates, would promote different stripe phases 
than those observed in the nickelates.

\begin{acknowledgments}
We thank K. Ro\'sciszewski for insightful discussions.
M.R. acknowledges support from the 
European Community under Marie Curie Program number HPMT2000-141. 
This work was supported by the Polish Ministry of Science and 
Education under Project No. 1~P03B~068~26, and by the Minist\`ere 
Fran\c{c}ais des Affaires Etrang\`eres under POLONIUM 09294VH.
\end{acknowledgments}



\begin{thebibliography}{999}
\bibitem{noteeg} The electronic structure in the three-dimensional case
                 was recently analyzed in Ref. \onlinecite{Fei05}.
\end{thebibliography}
\end{document}